\documentclass[10pt,a4paper]{article}
\usepackage{amsmath}
\usepackage{amssymb}
\usepackage{amsfonts}
\usepackage{amstext}
\usepackage{amsbsy}
\usepackage[mathscr]{eucal}
\usepackage{graphicx}
\usepackage{color}
\usepackage[all]{xy}
\newcommand{\beq}{\begin{equation}}
\newcommand{\eeq}{\end{equation}}
\newcommand{\beqa}{\begin{eqnarray}}
\newcommand{\eeqa}{\end{eqnarray}}
\newcommand{\ket}[1]{| #1 \rangle}

\newtheorem{thm}{Theorem}[subsection]

 \newtheorem{prop}[thm]{Proposition}
\newtheorem{defn}[thm]{Definition}

 \numberwithin{equation}{subsection}
\makeindex
\title{\Large\textbf{Algebraic structures of multipartite quantum systems}}

\author{\textit{ Hoshang Heydari}\\
        \small\textit{Institute of Quantum
Science, Nihon University,}\\
\small\textit{1-8 Kanda-Surugadai, Chiyoda-ku, Tokyo 101-8308, Japan
}}

\date{}
\begin{document}
\maketitle \thispagestyle{empty} \maketitle %%%%%%%%%%%%%%%%%

\begin{abstract}
We investigate the relation between multilinear mappings and
multipartite states.  We show that the isomorphism between
multilinear mapping and tensor product completely characterizes
decomposable multipartite states in a mathematically well-defined
manner.
\end{abstract}
\section{Introduction}
The characterization of multipartite state is a fascinating
subject in the field of fundamental quantum theory with
interesting applications in quantum information and quantum
computing. There are
 many multipartite states which can be used for different algorithm or
scheme for quantum computation. For example entangled cluster states
are the building block for one-way quantum computer as scheme for
universal quantum computation.  In recent years, we have witnessed
some progress in quantification and classification of multipartite
states, but this problem is still open and needs further
investigation.

In this paper we will discuss the structure of multipartite
product state in a clear and abstract algebraic way. Our main
interest are multilinear mapping or $m$-linear mapping of complex
vector spaces. We will establish an isomorphism between this maps
and tensor product states. In section \ref{bilmb} we will give an
introduction to the structure of bilinear mapping and condition
for which we have isomorphic mapping between this bilinear mapping
and tensor product. We will also show that this construction
defines the separable set of general bipartite states. Moreover,
we will establish a relation between this construction and the
concurrence. In section \ref{bilmm}, we will generalize our result
from bilinear to multilinear or $m$-linear mapping and the
construction of tensor product for such mapping. We will show that
this construction describes the separable set of general
multipartite states. Finally, we will establish a relation between
this  algebraic construction and the generalized concurrence. An
introduction to theory of multilinear mapping and algebra can be
found in \cite{Greub1} which is also our main reference.

\section{Bilinear mapping and bipartite state}\label{bilmb}
In this section we will give an introduction to bilinear mapping and
tensor product.  Let $\mathrm{V}_{1},\mathrm{V}_{2}$ be two linear
spaces and consider the  mapping
\begin{equation}\label{mlinear}
\Phi:\mathcal{V}_{1}\times\mathcal{V}_{2}\longrightarrow\mathcal{M}
\end{equation}
This map is called bilinear if it satisfies the following
condistions\begin{enumerate}
             \item $\Phi(\lambda \ket{\psi^{1}_{1}}+\mu \ket{\psi^{2}_{1}},\ket{\psi_{2}})=
             \lambda\Phi( \ket{\psi^{1}_{1}},\ket{\psi_{2}}))
             +\mu\Phi( \ket{\psi^{2}_{1}},\ket{\psi_{2}})$ for all
             $\ket{\psi^{1}_{1}},\ket{\psi^{2}_{1}}\in\mathcal{V}_{1}$,
             $\ket{\psi_{2}}\in\mathcal{V}_{2}$, and $\lambda, \mu \in\mathcal{N}$
             \item $\Phi(\ket{\psi_{1}},\lambda \ket{\psi^{1}_{2}}+\mu \ket{\psi^{2}_{2}})=
             \lambda\Phi( \ket{\psi_{1}},\ket{\psi^{1}_{2}}))
             +\mu\Phi( \ket{\psi_{1}},\ket{\psi^{2}_{2}})$ for all
             $\ket{\psi^{1}_{2}},\ket{\psi^{2}_{2}}\in\mathcal{V}_{2}$,
             $\ket{\psi_{1}}\in\mathcal{V}_{1}$,
           \end{enumerate}
for a linear space $\mathcal{N}$. We call this bilinear algebra a
bilinear function if the linear space $\mathcal{N}=\mathcal{M}$. In
this paper, we are mostly interested in complex vector spaces and
specially finite dimensional. The most important observation about
bilinear mapping is that the set of all vectors $\mathcal{S}$ in
$\mathcal{M}$ of the form $\Phi( \ket{\psi_{1}},\ket{\psi_{2}})$ for
$\ket{\psi_{1}}\in\mathcal{V}_{1}$ and
$\ket{\psi_{2}}\in\mathcal{V}_{2}$ is not in general a linear
subspace of the target space $\mathcal{M}$. To make this important
point clear we will give an example based on pair qubits. In this
case we have $\mathcal{V}_{1}=\mathcal{V}_{2}=\mathbb{C}^{2}$ and
$\mathcal{M}=\mathbb{C}^{4}$. Let also
$\ket{\psi_{1}}=\alpha^{1}_{1}\ket{e_{1}}+\alpha^{1}_{2}\ket{e_{1}}$
and
$\ket{\psi_{2}}=\alpha^{2}_{1}\ket{e_{1}}+\alpha^{2}_{2}\ket{e_{1}}$
In this case the bilinear mapping $\Phi(
\ket{\psi_{1}},\ket{\psi_{2}})$ is defined by
\begin{equation}
\Phi(\ket{\psi_{1}},\ket{\psi_{2}})=\alpha^{1}_{1}\alpha^{2}_{1}\ket{f_{1}}
+\alpha^{1}_{1}\alpha^{2}_{2}\ket{f_{2}}+\alpha^{1}_{2}\alpha^{2}_{1}\ket{f_{3}}
+\alpha^{1}_{2}\alpha^{2}_{2}\ket{f_{4}},
\end{equation}
where $\ket{f_{j}}$ is a basis for $\mathcal{M}$. Then a vector
$\ket{\psi}=\sum_{j}\alpha^{j}\ket{f_{j}}\in\mathcal{M}$ is
contained in the set $\mathcal{S}$ if and only if the components
satisfy the following very important condition:
\begin{equation}
\alpha^{1}\alpha^{4}=\alpha^{2}\alpha^{3}.
\end{equation}
This condition is exactly the separability condition for pair of
qubits. Let us consider the a pure two qubit state
$\ket{\Psi}=\sum^{1}_{i_{1}=0}\sum^{1}_{i_{2}=0}
\alpha_{i_{1}i_{2}} \ket{i_{1}}\otimes\ket{i_{2}}$. Then for this
state the separability condition  is given by
$\alpha_{00}\alpha_{11}=\alpha_{01}\alpha_{10}$. This equation
also gives a well-known measure of entanglement called the
concurrence
$C(\ket{\Psi})=2|\alpha_{00}\alpha_{11}-\alpha_{01}\alpha_{10}|$
for a pair of qubit \cite{Wootters98}.

Thus this is very important to investigate the algebra of product
states for bipartite and multipartite states. In following
sections we will use the notation $\mathrm{Im}\Phi$ for the
subspace of $\mathcal{M}$ which is generated by the set
$\mathcal{S}$.
\begin{defn} Let
$\mathrm{V}_{1},\mathrm{V}_{2}$ be two complex vector spaces and
consider the mapping $
\Phi:\mathcal{V}_{1}\times\mathcal{V}_{2}\longrightarrow\mathcal{M}
$. Then the pair $(\Phi,\mathcal{M})$ is called a tensor product if
and only if the following important conditions are satisfied:
\begin{itemize}
  \item $\mathrm{I}$: the image of bilinear mapping is equal the target space $\mathrm{Im}\Phi=\mathcal{M}$
  \item $\mathrm{II}$: If there is bilinear mapping $
\Psi:\mathcal{V}_{1}\times\mathcal{V}_{2}\longrightarrow\mathcal{N}
$, where $\mathcal{N}$ is a arbitrary complex vector space, then
there exists a linear mapping
$\Theta\mathcal{M}\longrightarrow\mathcal{N}$ such that the
following diagram $$\xymatrix{\mathcal{V}_{1}\times\mathcal{V}_{2}
\ar[d]_{\Psi}\ar[r]_{\Phi}&\mathcal{M}
\ar[d]_{\Theta}\\
             \mathcal{N}\ar@{=}[r]&\mathcal{N}
             }$$
             is commutative, that is $\Psi=\Theta\circ\Phi$.
\end{itemize}
\end{defn}
If $(\Phi,\mathcal{M})$ is a tensor product, then we denote
$\mathcal{M}=\mathcal{V}_{1}\otimes\mathcal{V}_{2}$ and
$\Phi(\ket{\phi_{1}},\ket{\phi_{2}})=\ket{\phi_{1}}\otimes\ket{\phi_{2}}$.
Moreover, the bilinearity is give by
\begin{enumerate}
             \item $(\lambda \ket{\psi^{1}_{1}}+\mu \ket{\psi^{2}_{1}})\otimes\ket{\psi_{2}}=
             \lambda\ket{\psi^{1}_{1}}\otimes\ket{\psi_{2}}
             +\mu\ket{\psi^{2}_{1}}\otimes\ket{\psi_{2}}$ for all
             $\ket{\psi^{1}_{1}},\ket{\psi^{2}_{1}}\in\mathcal{V}_{1}$,
             $\ket{\psi_{2}}\in\mathcal{V}_{2}$, and $\lambda, \mu \in\mathcal{N}$
             \item $\ket{\psi_{1}}\otimes(\lambda \ket{\psi^{1}_{2}}+\mu \ket{\psi^{2}_{2}})=
             \lambda\ket{\psi_{1}}\otimes\ket{\psi^{1}_{2}}))
             +\mu \ket{\psi_{1}}\otimes\ket{\psi^{2}_{2}}$ for all
             $\ket{\psi^{1}_{2}},\ket{\psi^{2}_{2}}\in\mathcal{V}_{2}$,
             $\ket{\psi_{1}}\in\mathcal{V}_{1}$.
           \end{enumerate}
Next, we will give some elementary properties of the tensor product.
For example every vector
$0\neq\ket{\psi}\in\mathcal{V}_{1}\otimes\mathcal{V}_{2}$ can be
written as
$\ket{\psi}=\sum^{k}_{i=1}\ket{\psi^{i}_{1}}\otimes\ket{\psi^{i}_{2}}$
for linearly independent vectors $\ket{\psi^{i}_{1}}$ and
$\ket{\psi^{i}_{2}}$. To see that let us choose a representation of
$\ket{\psi}$ such that it minimize the $k$. For $k=1$, it follows
easily that $\ket{\psi^{1}_{1}}\neq0$ and $\ket{\psi^{1}_{2}}\neq0$
and so $\ket{\psi^{1}_{1}}$ and $\ket{\psi^{1}_{2}}$ are linearly
independent vectors. Next, we show that the case for $k\geq 2$ is
also correct. For a linearly dependent vector we can assume that
$\ket{\psi^{i}_{1}}=\sum^{k-1}_{i=1}\gamma^{i}\ket{\psi^{i}_{1}}$,
so we have
\begin{eqnarray}
\ket{\psi}&=&\sum^{k-1}_{i=1}\ket{\psi^{i}_{1}}\otimes\ket{\psi^{i}_{2}}
+\sum^{k-1}_{i=1}\gamma^{i}\ket{\psi^{i}_{1}}\otimes\ket{\psi^{k}_{2}}
\\\nonumber&=&\sum^{k-1}_{i=1}\ket{\psi^{i}_{1}}\otimes(\ket{\psi^{i}_{2}}+\gamma^{i}\ket{\psi^{k}_{2}}=
\sum^{k-1}_{i=1}\ket{\psi^{i}_{1}}\otimes\ket{\psi^{i'}_{2}}.
\end{eqnarray}
This show that $k$ is not minimal. We can also show that
$\ket{\psi^{i}_{2}}$ are linearly independent in the same way as
above.

Now, let Let $\mathcal{V}_{1},\mathcal{V}_{2}$ be two complex vector
spaces and $\mathcal{V}_{1}\otimes\mathcal{V}_{2}$ be a tensor
product for these spaces. Moreover, let
$\mathcal{L}(\mathcal{V}_{1}\otimes\mathcal{V}_{2};\mathcal{M})$
denotes linear mapping
$\mathcal{V}_{1}\otimes\mathcal{V}_{2}\longrightarrow\mathcal{M}$
and $\mathcal{B}(\mathcal{V}_{1},\mathcal{V}_{2};\mathcal{M})$
denotes bilinear mapping
$\mathcal{V}_{1}\times\mathcal{V}_{2}\longrightarrow\mathcal{M}$.
Then, we have following isomorphism
\begin{equation}
\mathcal{L}(\mathcal{V}_{1}\otimes\mathcal{V}_{2};\mathcal{M})\longrightarrow
\mathcal{B}(\mathcal{V}_{1},\mathcal{V}_{2};\mathcal{M})
\end{equation}
which is defined by $\Phi(\Theta)=\Theta\circ\otimes$ for all
$\Theta\in\mathcal{L}(\mathcal{V}_{1}\otimes\mathcal{V}_{2};\mathcal{M})$.
The proof follows from conditions I and II for tensor product.
Moreover, the correspondence between the linear map
$\Psi\in\mathcal{L}(\mathcal{V}_{1}\otimes\mathcal{V}_{2};\mathcal{M})$
and $
\Theta\in\mathcal{B}(\mathcal{V}_{1},\mathcal{V}_{2};\mathcal{M})$
is visualize in following commutative diagram
\begin{equation}\xymatrix{\mathcal{V}_{1}\times\mathcal{V}_{2}
\ar[d]_{\otimes}\ar[r]_{\Phi}&\mathcal{M}
\ar@{=}[d]\\
             \mathcal{V}_{1}\otimes\mathcal{V}_{2}\ar[r]_{\Theta}&\mathcal{M}
             }
\end{equation}
Thus we have the following proposition
\begin{prop}
Let
$\Psi\in\mathcal{L}(\mathcal{V}_{1}\otimes\mathcal{V}_{2};\mathcal{M})$
be a bilinear map and $
\Theta\in\mathcal{B}(\mathcal{V}_{1},\mathcal{V}_{2};\mathcal{M})$
be induced linear map. Then $\Theta$ is surjective and injective if
and only if $\Phi$ satisfies the condition $\mathrm{I}$ and
$\mathrm{II}$ of tensor respectively.
\end{prop}
The map $\Theta$ is surjective follows from
$\mathrm{Im}\Psi=\mathrm{Im}\Theta$. Now, if we assume that $\Theta$
is injective, then $(\mathrm{Im}\Psi,\Psi)$ is a tensor product for
$\mathcal{V}_{1}$ and $\mathcal{V}_{2}$ and every bilinear mapping
$\mathcal{V}_{1}\times\mathcal{V}_{2}\longrightarrow\mathcal{N}$
induces a linear mapping
$\Upsilon:\mathrm{Im}\Psi\longrightarrow\mathcal{N}$ such that
$\Phi(\ket{\psi_{1}},\ket{\psi_{2}})=\Upsilon\Psi(\ket{\psi_{1}},\ket{\psi_{2}})$.
Next, $\Psi$ satisfies the condition $\mathrm{II}$ since if $\Theta$
is an extension of $\Upsilon$ to a map
$\Theta:\mathcal{M}\longrightarrow\mathcal{N}$, then
$\Phi(\ket{\psi_{1}},\ket{\psi_{2}})=\Theta\Psi(\ket{\psi_{1}},\ket{\psi_{2}})$.
The converse follows by assuming that $\Psi$ satisfies the condition
$\mathrm{II}$ and show that $\Theta$ is injective.

As an example let us look at the general bipartite states. For such
system we have a bilinear mapping
$\Sigma:\mathbb{C}^{N_{1}}\times\mathbb{C}^{N_{2}}\longrightarrow\mathrm{M}^{N_{1}\times
N_{2}}$ defined by
\begin{equation}\label{matrixbi}
(\alpha^{1}_{1},\alpha^{2}_{1},\ldots,\alpha^{N_{1}}_{1})\times
(\alpha^{1}_{2},\alpha^{2}_{2},\ldots,\alpha^{N_{2}}_{2})\longrightarrow
\left(%
\begin{array}{cccc}
 \alpha^{1}_{1}\alpha^{1}_{2}& \alpha^{1}_{1}\alpha^{2}_{2} & \cdots &\alpha^{1}_{1}\alpha^{N_{2}}_{2} \\
  \alpha^{2}_{1}\alpha^{1}_{2}& \alpha^{2}_{1}\alpha^{2}_{2} & \cdots &\alpha^{2}_{1}\alpha^{N_{2}}_{2} \\
  \vdots & \vdots&\ddots &\vdots \\
   \alpha^{N_{1}}_{1}\alpha^{1}_{2}& \alpha^{N_{1}}_{1}\alpha^{2}_{2} & \cdots &\alpha^{N_{1}}_{1}\alpha^{N_{2}}_{2} \\
\end{array}%
\right).
\end{equation}
But for this bilinear mapping the pair $(\mathrm{M}^{N_{1}\times
N_{2}},\Sigma)$ is a tensor product of $\mathbb{C}^{N_{1}}$ and
$\mathbb{C}^{N_{2}}$, that is
\begin{equation}\xymatrix{\mathbb{C}^{N_{1}}\times\mathbb{C}^{N_{2}}
\ar[d]_{\otimes}\ar[r]_{\Sigma}&\mathrm{M}^{N_{1}\times N_{2}}
\ar@{=}[d]\\
             \mathbb{C}^{N_{1}}\otimes\mathbb{C}^{N_{1}}\ar[r]_{\Theta}&\mathrm{M}^{N_{1}\times
N_{2}}
             }
\end{equation}
and thus represent the product states of general bipartite states.
 Let us consider the a general pure bipartite state
$\ket{\Psi}=\sum^{N_{1}-1}_{i_{1}=0}\sum^{N_{2}-1}_{i_{2}=0}
\alpha_{i_{1}i_{2}} \ket{i_{1}}\otimes\ket{i_{2}}$ and
$\mathrm{M}^{N_{1}\times
N_{2}}=\{\alpha_{i_{1}i_{2}}\in\mathbb{C}^{N_{1}}\times\mathbb{C}^{N_{2}}:
\alpha_{k_{1}k_{2}}\alpha_{l_{1}l_{2}}=\alpha_{l_{1}k_{2}}\alpha_{k_{1}l_{2}},
\forall i_{j}=k_{j},l_{j}, j=1,2\}$. Then for this state the
separability condition is give by
$\alpha_{k_{1}k_{2}}\alpha_{l_{1}l_{2}}=\alpha_{l_{1}k_{2}}\alpha_{k_{1}l_{2}}$.
This equation also gives a general expression for the concurrence
\begin{equation}
C(\ket{\Psi})=\left(\mathcal{N}\sum^{N_{1}-1}_{l_{1}>k_{1}=0}\sum^{N_{2}-1}_{l_{2}>k_{2}=0}|\alpha_{k_{1}k_{2}}\alpha_{l_{1}l_{2}}
-\alpha_{l_{1}k_{2}}\alpha_{k_{1}l_{2}}|^{2}\right)^{1/2}
\end{equation}
of a general bipartite state \cite{Hosh4}. Now, we will discuss
the direct decompositions which is the one interesting property of
tensor product. Let $\mathcal{V}_{1}$ and $\mathcal{V}_{2}$ be two
complex vector spaces and there is a direct decompositions of
these space as $\mathcal{V}_{1}=\sum_{r}\mathcal{V}^{r}_{1}$ and
$\mathcal{V}_{2}=\sum_{s}\mathcal{V}^{s}_{2}$. Moreover, assume
that the pair $(\mathcal{V}_{1}\otimes\mathcal{V}_{2},\otimes)$ is
a tensor product of these spaces. Then
$\mathcal{V}_{1}\otimes\mathcal{V}_{2}$ is the direct sum of the
subspaces $\mathcal{V}^{r}_{1}\otimes\mathcal{V}^{s}_{2}$, that is
$\mathcal{V}_{1}\otimes\mathcal{V}_{2}=\sum_{r}\sum_{s}\mathcal{V}^{r}_{1}\otimes\mathcal{V}^{s}_{2}$.
The first condition $\mathrm{I}$ follows from the observation that
$\mathcal{V}_{1}\otimes\mathcal{V}_{2}$ is generated by
$\ket{\psi_{1}}\otimes\ket{\psi_{2}}$ for
$\ket{\psi_{j}}\in\mathcal{V}_{j}$, $j=1,2$. But
$\ket{\psi_{1}}\otimes\ket{\psi_{2}}=\sum_{r}\sum_{s}\ket{\psi^{r}_{1}}\otimes\ket{\psi^{s}_{2}}$
for $\ket{\psi_{1}}=\sum_{r}\ket{\psi^{r}_{1}}$ and
$\ket{\psi_{2}}=\sum_{s}\ket{\psi^{s}_{2}}$ with
$\ket{\psi^{r}_{1}}\in\mathcal{V}^{r}_{1}$ and
$\ket{\psi^{s}_{2}}\in\mathcal{V}^{s}_{2}$. Thus
$\mathcal{V}_{1}\otimes\mathcal{V}_{2}$ is the sum of the
subspaces $\mathcal{V}^{r}_{1}\otimes\mathcal{V}^{s}_{2}$. It is
more difficult to show that the decomposition is direct  and the
proof can be found in \cite{Greub1}.

%%%%%%%%%%%%%%%%%%%%%%
\section{Multilinear mapping and multipartite states}\label{bilmm}
In this section we will give an introduction to multilinear mapping
and tensor product. The relation between the multilinear mapping and
tensor product gives the product states of multipartite states.  Let
$\mathcal{V}_{1},\mathcal{V}_{2},\ldots,\mathcal{V}_{m}$ be $m$
complex vector spaces. Then the mapping
$\Phi:\mathcal{V}_{1}\times\mathcal{V}_{2}\times\cdots\times\mathcal{V}_{m}\longrightarrow\mathcal{M}$
is called $m$-linear if for every $j$ with $1\leq j\leq m$ we have
\begin{eqnarray}
&&\Psi(\ket{\psi_{1}},\ldots,\ket{\psi_{j-1}},\lambda\ket{\psi_{j}}+\mu\ket{\phi_{j}},\ket{\psi_{j+1}},
\ldots,\ket{\psi_{m}}\\\nonumber&&=\lambda\Psi(\ket{\psi_{1}},\ldots,\ket{\psi_{j-1}},\ket{\psi_{j}},\ket{\psi_{j+1}},
\ldots,\ket{\psi_{m}}\\\nonumber&&+\mu\Psi(\ket{\psi_{1}},\ldots,\ket{\psi_{j-1}},\ket{\phi_{j}},\ket{\psi_{j+1}},
\ldots,\ket{\psi_{m}},
\end{eqnarray}
where $\ket{\psi_{j}},\ket{\phi_{j}}\in\mathcal{V}_{j}$ and
$\lambda,\mu\in\mathcal{K}$. We also denote the image of the mapping
$\Psi$ by $\mathrm{Im}\Psi$,that is, the subspace of $\mathcal{M}$
which is generated by the vectors
$\Psi(\ket{\psi_{1}},\ket{\psi_{2}},\ldots,\ket{\psi_{m}}$.
Moreover, let
$\mathcal{L}(\mathcal{V}_{1},\mathcal{V}_{2},\ldots,\mathcal{V}_{m};\mathcal{M})$
be the set of all $m$-linear maps
$\mathcal{L}(\mathcal{V}_{1},\mathcal{V}_{2},\ldots,\mathcal{V}_{m}\longrightarrow\mathcal{M})$.
Then we can obtain a linear structure in
$\mathcal{L}(\mathcal{V}_{1},\mathcal{V}_{2},\ldots,\mathcal{V}_{m};\mathcal{M})$
by defining the following operations
\begin{equation}(\Psi+\Phi)(\ket{\psi_{1}},\ket{\psi_{2}},\ldots,\ket{\psi_{m}})=
\Psi(\ket{\psi_{1}},\ket{\psi_{2}},\ldots,\ket{\psi_{m}})+\Phi(\ket{\psi_{1}},\ket{\psi_{2}},\ldots,\ket{\psi_{m}})
\end{equation}
and
$(\lambda\Psi)(\ket{\psi_{1}},\ket{\psi_{2}},\ldots,\ket{\psi_{m}})=
\lambda\Psi(\ket{\psi_{1}},\ket{\psi_{2}},\ldots,\ket{\psi_{m}})$.
%%%%%%%%%%%%%
\begin{defn} Let
$\mathcal{V}_{1},\mathcal{V}_{2},\ldots,\mathcal{V}_{m}$ be $m$
complex vector spaces and consider he $m$-linear mapping
\begin{equation}\label{mlinear}
\Phi:\mathcal{V}_{1}\times\mathcal{V}_{2}\times\cdots\times\mathcal{V}_{m}\longrightarrow\mathcal{M}.
\end{equation} Then the pair $(\Phi,\mathcal{M})$ is called a tensor product if
and only if the following important conditions are satisfied:
\begin{itemize}
  \item $\mathrm{I}_{\otimes}$: the image of bilinear mapping is equal the target space $\mathrm{Im}\Phi=\mathcal{M}$
  \item $\mathrm{II}_{\otimes}$: If there is bilinear mapping $
\Psi:\mathcal{V}_{1}\times\mathcal{V}_{2}\times\cdots\times\mathcal{V}_{m}\longrightarrow\mathcal{N}
$, where $\mathcal{N}$ is a arbitrary complex vector space, then
there exists a linear mapping
$\Theta\mathcal{M}\longrightarrow\mathcal{N}$ such that the
following diagram
$$\xymatrix{\mathcal{V}_{1}\times\mathcal{V}_{2}\times\cdots\times\mathcal{V}_{m}
\ar[d]_{\Psi}\ar[r]_{\Phi}&\mathcal{M}
\ar[d]_{\Theta}\\
             \mathcal{N}\ar@{=}[r]&\mathcal{N}
             }$$
             is commutative, that is $\Psi=\Theta\circ\Phi$.
\end{itemize}
\end{defn}
We can denote the tensor product $(\Phi,\mathcal{M})$ of spaces
$\mathcal{V}_{j}$ by
$(\mathcal{V}_{1}\otimes\mathcal{V}_{2}\otimes\cdots\otimes\mathcal{V}_{m},\bigotimes^{m})$
and $\Phi(\ket{\phi_{1}},\ket{\phi_{2}},\ldots,\ket{\phi_{m}}
)=\ket{\phi_{1}}\otimes\ket{\phi_{2}}\otimes\cdots\otimes\ket{\phi_{m}}$.
%%%%%%%%%%%%%%
Now, let $\mathcal{V}_{1},\mathcal{V}_{2},\ldots,\mathcal{V}_{m}$ be
$m$ complex vector spaces and
$\mathcal{V}_{1}\otimes\mathcal{V}_{2}\otimes\cdots\otimes\mathcal{V}_{m}$
be a tensor product for these spaces. Moreover, let
$\mathcal{L}(\mathcal{V}_{1}\otimes\mathcal{V}_{2}\otimes\cdots\otimes\mathcal{V}_{m};\mathcal{M})$
denotes linear mapping
$\mathcal{V}_{1}\otimes\mathcal{V}_{2}\otimes\cdots\otimes\mathcal{V}_{m}\longrightarrow\mathcal{M}$
and
$\mathcal{L}(\mathcal{V}_{1},\mathcal{V}_{2},\ldots,\mathcal{V}_{m};\mathcal{M})$
denotes multilinear mapping
$\mathcal{V}_{1}\times\mathcal{V}_{2}\times\cdots\times\mathcal{V}_{m}\longrightarrow\mathcal{M}$.
Then, we have following isomorphism
\begin{equation}
\mathcal{L}(\mathcal{V}_{1}\otimes\mathcal{V}_{2}\otimes\cdots\otimes\mathcal{V}_{m};\mathcal{M})\longrightarrow
\mathcal{L}(\mathcal{V}_{1},\mathcal{V}_{2},\ldots,\mathcal{V}_{m};\mathcal{M})
\end{equation}
which is defined by $\Phi(\Theta)=\Theta\circ\otimes$ for all
$\Theta\in\mathcal{L}(\mathcal{V}_{1}\otimes\mathcal{V}_{2};\mathcal{M})$.
The proof follows from conditions $\mathrm{I}_{\otimes}$ and
$\mathrm{II}_{\otimes}$ for tensor product. Moreover, the
correspondence between the linear map
$\Psi\in\mathcal{L}(\mathcal{V}_{1}\otimes\mathcal{V}_{2}\otimes\cdots\otimes\mathcal{V}_{m};\mathcal{M})$
and $
\Theta\in\mathcal{B}(\mathcal{V}_{1},\mathcal{V}_{2},\ldots,\mathcal{V}_{m};\mathcal{M})$
is visualize in following commutative diagram
\begin{equation}\xymatrix{\mathcal{V}_{1}\times\mathcal{V}_{2}\times\cdots\times\mathcal{V}_{m}
\ar[d]_{\otimes}\ar[r]_{~~~~\Phi}&\mathcal{M}
\ar@{=}[d]\\
             \mathcal{V}_{1}\otimes\mathcal{V}_{2}\otimes\cdots\otimes\mathcal{V}_{m}\ar[r]_{~~~~\Theta}&\mathcal{M}
             }
\end{equation}
Thus in general case we have the following proposition
\begin{prop}
Let
$\Psi\in\mathcal{L}(\mathcal{V}_{1}\otimes\mathcal{V}_{2}\otimes\cdots\otimes\mathcal{V}_{m};\mathcal{M})$
be a mulilinear map and $
\Theta\in\mathcal{L}(\mathcal{V}_{1},\mathcal{V}_{2},\ldots,\mathcal{V}_{m};\mathcal{M})$
be induced linear map. Then $\Theta$ is surjective and injective if
and only if $\Phi$ satisfies the condition $\mathrm{I}_{\otimes}$
and $\mathrm{II}_{\otimes}$ of tensor respectively.
\end{prop}
Let $\mathcal{V}_{1},\mathcal{V}_{2},\ldots,\mathcal{V}_{m}$ be
complex vector spaces. Then the mapping
\begin{equation}
\Psi:\mathcal{L}(\mathcal{V}_{1})\times\mathcal{L}(\mathcal{V}_{2})\times\cdots\times\mathcal{L}(
\mathcal{V}_{m})\longrightarrow\mathcal{L}(\mathcal{V}_{1}\otimes\mathcal{V}_{2}\otimes\cdots\otimes\mathcal{V}_{m})
\end{equation}
given by
$\Psi(\Theta_{1},\ldots,\Theta_{m})(\ket{\psi_{1}}\otimes\cdots\otimes\ket{\psi_{m}})=\Theta_{1}(\ket{\psi_{1}})
\cdots\Theta_{m}(\ket{\psi_{m}})$ is a tensor product for the space
$\mathcal{L}(\mathcal{V}_{j})$.
%%%%%%%%%%%%

We will give an example to visualize the relation between our
$m$-linear construction and multipartite product states. Let us
consider a multipartite states where
$\mathcal{V}_{j}=\mathbb{C}^{N_{j}}$ for all $1\leq j\leq m$ and a
general state is given by
\begin{equation}
\ket{\Psi}=\sum^{N_{1}-1,N_{2}-1,\ldots,N_{m}-1}_{i_{1},i_{2},\ldots,i_{m}=0}
\alpha_{i_{1}i_{2}\cdots i_{m}}\ket{i_{1}}
\otimes\ket{i_{2}}\otimes\cdots\ket{i_{m}}.
\end{equation} Then we have the
following commutative diagram
\begin{equation}\xymatrix{\mathbb{C}^{N_{1}}\times\mathbb{C}^{N_{2}}\times\cdots\times\mathbb{C}^{N_{m}}
\ar[d]_{\otimes}\ar[r]_{~~~~\Phi}&\mathrm{M}^{N_{1}\times{N}_{2}\times\cdots\times{N}_{m}}
\ar@{=}[d]\\
             \mathbb{C}^{N_{1}}\otimes\mathbb{C}^{N_{2}}\otimes\cdots\otimes\mathbb{C}^{N_{m}}\ar[r]_{~~~~\Theta}&
             \mathrm{M}^{N_{1}\times{N}_{2}\times\cdots\times{N}_{m}}
             }
\end{equation}
where
$\mathrm{M}^{N_{1}\times{N}_{2}\times\cdots\times{N}_{m}}=\left(\alpha_{i_{1}i_{2}\cdots
i_{m}}\right)_{1\leq i_{j}\leq N_{j}}$, is a multi-box matrix
which is defined as follows \begin{equation}
\mathrm{M}^{N_{1}\times\cdots \times N_{m}}=\{\alpha_{i_{1}\ldots
i_{m}}\in\mathbb{C}^{N_{1}}\times\cdots\times\mathbb{C}^{N_{m}}:\mathcal{S}^{k_{j}l_{j}}_{1\leq
j\leq m}=0, \forall , j=1,2,\ldots, m\},
\end{equation}
where
\begin{eqnarray}\label{eq: submeasure}\mathcal{S}^{k_{j}l_{j}}_{1\leq j\leq m}
&=&\alpha_{k_{1}k_{2}\cdots k_{m}}\alpha_{l_{1}l_{2}\cdots
l_{m}}-\\\nonumber&& \alpha_{k_{1}k_{2}\cdots
k_{j-1}l_{j}k_{j+1}\cdots k_{m}}\alpha_{l_{1}l_{2}\cdots
l_{j-1}k_{j}l_{j+1}\cdots l_{m}}
\end{eqnarray}
This construction also gives a general expression for the
concurrence
\begin{equation}
C(\ket{\Psi})=\left(\mathcal{N}\sum^{N_{1}-1}_{l_{1}>k_{1}=0}\cdots
\sum^{N_{3}-1}_{l_{3}>k_{3}=0}|\mathcal{S}^{k_{j}l_{j}}_{1\leq
j\leq 3}|^{2}\right)^{1/2}
\end{equation}
of a general three-partite state. Note also that for a mixed state
a measure of entanglement can be constructed by taking the infimum
over all pure decomposition of a given state using above
expression for concurrence.
 This is our main result for
general multipartite states. This result is also related to the
construction of  Segre variety given in \cite{Hosh4}. We can also
construct a measure of entanglement for general multipartite
states based on the multidimensional matrix
$\mathrm{M}^{N_{1}\times{N}_{2}\times\cdots\times{N}_{m}}$ with
some additional structures \cite{Hosh6}.

\begin{flushleft}
\textbf{Acknowledgments:} The  author acknowledges the financial
support of the Japan Society for the Promotion of Science (JSPS).
\end{flushleft}

%%%%%%%%%%%%%%%%%%%%%%%%%%%%%%%%%%%%%%%%%%%%%%%%%%%%%%%%%%%%%%%%%%

\end{document}